\documentclass[manuscript]{aastex}
\usepackage[english]{babel}
\usepackage{lipsum}
\usepackage{graphicx}
\usepackage{color}
\usepackage[pagewise]{lineno}

\begin{document}

\title{Dynamic Precursors of Flares in Active Region NOAA 10486 }

\author{M. B. Kors\'os\altaffilmark{1,2}, N. Gyenge\altaffilmark{1,2}, T. Baranyi\altaffilmark{1} and A. Ludm\'any\altaffilmark{1}}
\altaffiltext{1}{Debrecen Heliophysical Observatory (DHO), Research Centre for Astronomy and Earth Sciences, Hungarian Academy of Science, 4010 Debrecen, P.O. Box 30, Hungary}
\altaffiltext{2}{Solar Physics \& Space Plasma Research Center (SP2RC), University of Sheffield, Hounsfield Road, S3 7RH, UK}
\email{[korsos.marianna; gyenge.norbert; baranyi.tunde; ludmany.andras]@csfk.mta.hu}

\begin{abstract}
Four different methods are applied here to study the precursors of flare activity in the Active Region NOAA 10486. Two approaches track the temporal behaviour of suitably chosen features (one, the weighted horizontal gradient $WG_{M}$, is generalised form the horizontal gradient of the magnetic field, $G_{M}$; another is the sum of  the horizontal gradient of the magnetic field, $G_{S}$, for all sunspot pairs). $WG_{M}$ is a photospheric indicator that is a proxy measure of magnetic non-potentiality of a specific area of the active region, i.e.  it captures the temporal variation of the weighted horizontal gradient of magnetic flux summed up for the region where opposite magnetic polarities are highly mixed. 
The third one, referred to as the separateness parameter, $S_{l-f}$, considers the overall morphology. 
Further, $G_{S}$ and $S_{l-f}$ are photospheric newly defined quick-look indicators of the polarity mix of the entire active region. The fourth method is tracking the temporal variation of small x-ray flares, their times of succession and their energies observed by the Reuven Ramaty High Energy Solar Spectroscopic Imager instrument.  All approaches yield specific pre-cursory signatures for the imminence of flares.
\end{abstract}


\section{Introduction}
The main question of Space Weather is how to optimise the forecast of flares and Coronal Mass Ejections (CMEs). Flares and CMEs are important ingredients of Space Weather in order to protect mankind and technological systems that might also be at risk from Space Weather effects. One needs to understand clearly the causes and dynamics of Space Weather phenomena.  After some significant developments in flare-forecast methods in the past decade the reliable prediction of such eruptive events in solar active regions remains  an important and challenging central task. It is not yet known all the details of the flare phenomenon, therefore, it is difficult to develop a well-functioning flare forecasting method. One usually searches for global features, that is capable to predict these impulsive events  in appropriate time. Several flare models claim that photospheric shear motions may be responsible for the energy build-up and flare eruption. Others examine parameters derived from measurements of the magnetic field. 

Let us here recall a few relevant studies to this paper. Ambastha (1993) investigated the evolution of the degree of magnetic shear in flare-producing active regions. Mathew and Ambastha (2000) suggested that "flux motions, cancellation, and large magnetic field gradient, may be the prerequisite criterion for the triggering of this flare". Wang (2006) examined the rapid gradient of the photospheric magnetic fields along the flaring neutral lines in five different ARs. Some other examples of potential flare indicators are, e.g., the pre-defined K-parameter describing the strength of the gradient across the neutral line \citep{Schrijver}; flux-normalized measure of the field twist \citep{falcon2002}; maximum horizontal gradient and the length of the neutral line \citep{cui2006}; and, shear \citep{cui2007}. Other groups focused on the spatial and temporal resolution of the magnetic helicity injection in the magnetic field, e.g. Vemareddy (2012) and Zhang (2008).

Once a pre-flare parameter is introduced, it has to be put to test, i.e., how effectively actually it works. Steward et. al. (2011) studied the length of the strong-gradient polarity inversion lines (SPILs) and tested the predictiability of x-ray flares of class C, M and X in a 24-h-interval exceeding 88 $\%$ of target accuracy. In a previous paper (Kors\'os {\it et al.}, 2014, hereafter Paper I) we presented a novel approach, where the temporal variation of a pre-defined parameter, the so-called horizontal magnetic gradient (denoted as $G_{M}$), was traced. The method of Paper I yields additional advantages in comparison with the static approaches. Here, this paper presents further new and reliable  of the pre-flare dynamics which are expected to be useful and practical diagnostic tools.

\section{Observational Data}

This paper investigates the Active Region NOAA 10486. The following datasets were used for the examination. The SOHO/MDI-Debrecen Data (also known as SDD\footnote{http://fenyi.solarobs.unideb.hu/SDD/SDD.html}) is the most detailed and precise sunspot database between 1996 and 2011. This catalogue contains information on the position, area and magnetic field of all observable sunspots and sunspot groups on a 1.5-hourly basis \citep{SDD}. This comprehensive database allows to follow the evolution of the magnetic configuration of the Active Region NOAA 10486 in high spatial and temporal resolution. 

The data of major solar flares were taken from the Geostationary Operational Environmental Satellite (also known as GOES) x-ray flare database. The method described in Paper I only allows to predict the data (e.g. time and intensity) of flares stronger than M5 according to the X-flare classification scheme of GOES in the 1-8 $\AA$ wavelength interval\footnote{https://www.nsof.class.noaa.gov/release/data/available/goes/index.htm}. The present examination is also restricted to these cases.

The third data source applied here is the Reuven Ramaty High Energy Solar Spectroscopic Imager (also known as RHESSI) database. Since its launch, the RHESSI \citep{Lin} satellite has observed more than 95,000 microflare events\footnote{http://hesperia.gsfc.nasa.gov/rhessi3/data-access/rhessi-data/rhessi-data/index.html} in 9 different energy channels. These events are displayed in a table consisting of the main parameters of flares: time of explosion, durations, peak intensities, total counts during the outburst, energy channel of the maximal energy at which the flare is still measurable, location on the solar disc and quality flags.

\section{Evolution of magnetic field of active regions}

\subsection{$G_{M}$ between nearby groups of spots of opposite magnetic polarities}

In Paper I, a new physical parameter was introduced that may characterise the evolution of the magnetic field in the flare-producing domain of an active region. The horizontal gradient of the magnetic field ($G_{M}$) between two spots of opposite magnetic polarities is:

\begin{equation}
	G_{M}= \left | \frac {f(A_{1})*A_{1} - f(A_{2})*A_{2}} {d} \right | ,
\end{equation}

\noindent
where {$A_{1}$ and {$A_{2}$ are the areas of two umbrae, $f(A_{1})$ and $f(A_{2})$ represent the mean flux density of two spots expressed as a function of the umbral area. The product of the two quantities, $f(A)$*$A$, is a good proxy for the flux amount in the umbra. The function $f(A)$ is determined by using the SDD data. This representation helps to avoid the problem of decreasing precision of the magnetic field towards the solar limb. Initially, the  $d$ is distance of the two spots.

The present work proposes a generalised form of the above formula (Equation 1). The weighted horizontal gradient $WG_{M}$ is now computed between two groups of nearby spots having opposite polarities, the flux amounts ($f(A_{p})$*$A_{p}$ and $f(A_{n})$*$A_{n}$) are summarized for the two groups and the distance ($d$) is computed between their centers of weight.
\begin{equation}
	WG_{M}= \left | \frac {f(A_{p})*A_{p} - f(A_{n})*A_{n}} {d} \right | .
\end{equation} 

 Similar to Paper I, this quantity is considered to be a potential proxy of magnetic non-potentiality at the photospheric level. Its variation is traced in the domain of the highest magnetic gradient.

 \begin{figure}
	\centerline{\includegraphics[height=0.187\textheight]{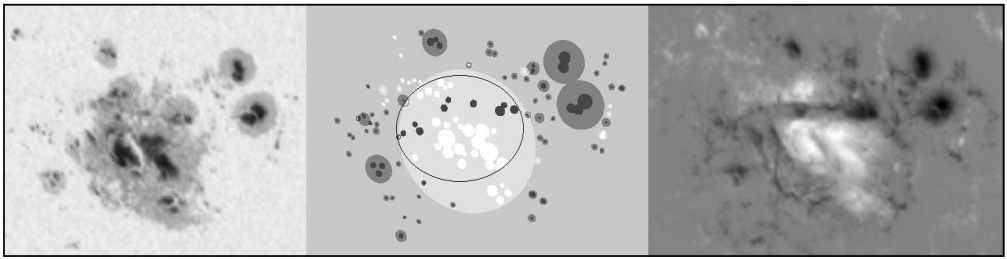}}
	\caption{Images: AR NOAA 10486 on 28 October 2003 at 01:35 UT: continuum image (left panel), reconstruction from SDD (middle panel), magnetogram (right panel).}
	\label{10486}
\end{figure}

Figure \ref{10486} shows the Active Region NOAA 10486, its white-light appearance (left panel), the view of the sunspot group reconstructed from the SDD data (middle panel) and the corresponding magnetogram (right panel). This Active Region produced numerous flares, however, only the events stronger than M5 are considered here. The cartoon of the active region in Figure \ref{10486} (middle panel) visualises the highlighted area containing spots of opposite polarities. The study area is where the most intense ßares are in connection with the location of the strongest magnetic gradient. The small black (negative polarity) and white (positive polarity) circles are visualisations of sunspots. The larger bright/dark-shaded circles are the corresponding penumbrae. For more details see Paper I.

 Figure \ref{10486grad} shows the temporal variation of the $WG_{M}$ (Figure \ref{10486grad}A), distance (Figure \ref{10486grad}B) and flux amount, i.e. the sum of unsigned flux of all spots (Figure \ref{10486grad}C) between 25 October and 3 November with a cadence of 1.5 hours.} It is clearly visible that a steep rise and a high maximum value of the weighted horizontal gradient is followed by a less steep decrease which ends up with an energetic X1.2 flare on 26 October and two medium flares with strengths M5.0 and M6.7 on 27 October. Then, two further similar $WG_{M}$ variations can be observed ending with an X17.2 event on 28 October and an X10 event on 29 October. \cite{deng} used high-resolution data and reported a decrease of magnetic field gradients  after the occurrence of an X10 flare, similar to Figure \ref{10486grad}A. Finally, during the period of investigation, a third flare event took place where the above-mentioned typical behaviour of the $WG_{ M}$ parameter ended up with an energetic X8.3 flare on 02 November.

Parallel to the increasing/decreasing trends of $WG_{M}$ decreasing/increasing trends of the distance can also be observed. This type of pre-flare behaviour was already reported in Paper I between spot pairs of opposite polarities chosen at the location of the highest magnetic gradients. The present results, however, unveil that this behaviour is also exhibited by sub-groups of opposite polarities in this region. This is a promising phenomenon for flare forecast.

 \begin{figure}
	\centerline{\includegraphics[height=0.58\textheight]{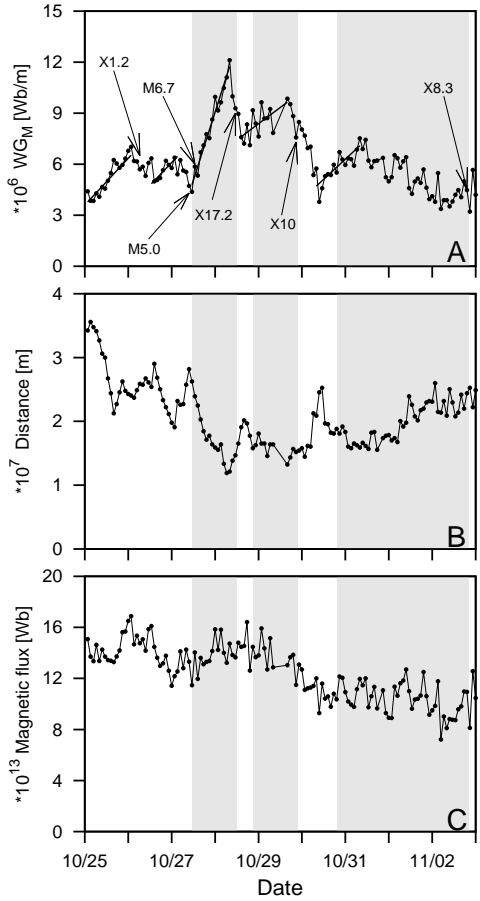}}
	\caption{Variations in Active Region NOAA 10486 within the highlighted area (see in the Fig. \ref{10486} middle panel) between 25 October and 3 November, 2003.  A: variations of $WG_{M}$ ; B: distance of opposite polarity regions; C: flux amount, the total unsigned flux of all spots. The gray stripes indicate those time intervals in which the pre-flares are examined, see also Figures \ref{precursors_X17} and Figure \ref{precursors_X10}.} 
	\label{10486grad}
\end{figure}

\subsection{Measures of the deviation from the bipolar structure}

Several attempts have been made in the past to classify the internal morphology of sunspot groups, see e.g. the Z\"urich-McIntosh (\citealt{McIntosh}) and Mount Wilson classification\footnote{http://www.spaceweather.com/glossary/magneticclasses.html} schemes. These were only suitable to an approximate assessment of the level of non-potentiality of solar active regions because they do not quantify information about magnetism. The SDD and SDO/HMI - Debrecen Data (HMIDD\footnote{http://fenyi.solarobs.unideb.hu/ESA/HMIDD.html}) catalogues provide the possibility to measure the level of separateness/mix by a single parameter which is suitable for scaling the deviation from the classic bipolar structure. Our aim is to introduce specific parameters that indicate which actual individual sunspot group should be given a priority to carry out a detailed real-time follow-up of $G_{M}$ using our method. The Debrecen sunspot catalogues (SDD and HMIDD) allow us to define numerical measures of the mixed states of sunspots with opposite polarities. We introduce and examine now two new parameters. 

 \begin{figure}
	\centerline{\includegraphics[height=0.58\textheight]{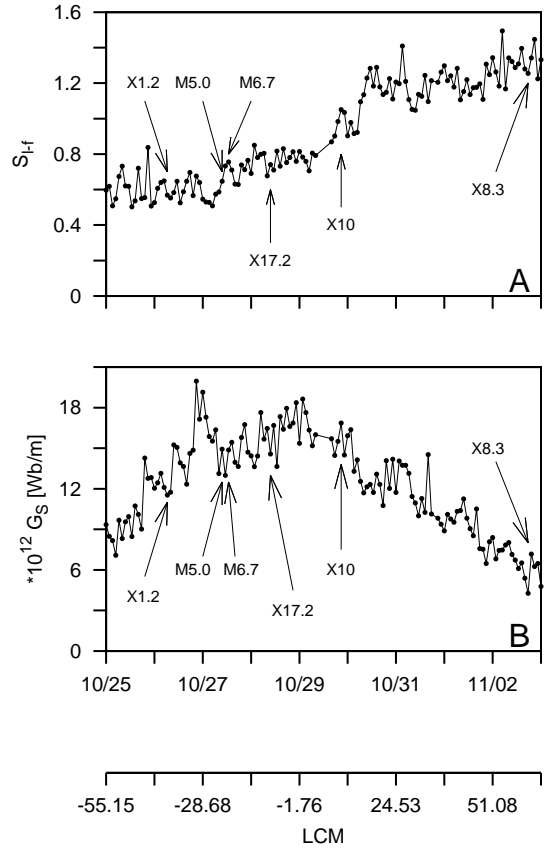}}
	\caption{Variations in Active Region NOAA 10486 within the highlighted area (see in the Fig.  \ref{10486}, middle panel) between 25 October and 3 November, 2003. A: pre-flare variation of $S_{l-f}$; B: pre-flare variation of $G_{S}$. The lower axis shows the LCM (Length of Central Meridian) of AR NOAA 10486}
	\label{10486parameter}
\end{figure}

The first parameter is based on the method described in Paper I, following the measure of the horizontal magnetic gradient of opposite polarity spots. Here, we generalise this parameter by introducing the sum of $G_{M}$ for all spot-pairs of opposite polarities within the group:

 \begin{equation}
	G_{S}=\left | \sum_{i,j} \frac{B_{p,i}A_{p,i}-B_{n,j}A_{n,j}}{d_{i,j}}\right |.
\end{equation}
  
The sum of $G_{M}$ for all spot-pairs of opposite polarities (i.e. $G_{S}$) within the group exhibits a similar temporal variation to the $G_{M}$: a steep rise and some decrease prior to flares, see panel B of Figure~\ref{10486parameter}. A high value of $G_{S}$ means that the group contains several spot-pairs with high horizontal magnetic gradient. Thus, both the temporal variation and the amplitude of this parameter are indicative for flare productivity, but this parameter is primarily a dynamic indicator of the flare risk.

The other complexity parameter is the so-called separateness parameter that may well characterise the mutual mixing of the opposite polarity subgroups, i.e.

\begin{equation}
	S_{l-f}=\frac{d_{lc-fc}}{2\sqrt{{\sum A_{i}/\pi}}}.
\end{equation}

The numerator is the distance of the centers of weight of the leading and following subgroups, the denominator is the diameter of a circle having an area equal to the total area of the umbrae. The smaller this ratio is the more mixed is the distribution of the spots of opposite polarities. If it is about unity or lower, then, the polarities are highly mixed and the active region shows flare risk, see panel A of Figure~\ref{10486parameter}. If, however, its value is found to be about four, then, the centers of the leading and following subgroups are separated (see \citealt{Korsosb}). In contrast to the $G_{S}$ parameter, this latter one is a static indicator and does not contain signatures of the intensity and time of an imminent flare, just its probability. It is advantageous that its determination can be more easily automated. 

The two parameters defined above are attempts to replace the traditional (Z\"urich, McIntosh, Mount Wilson) classification schemes by a more appropriate parameter based on data of sunspots and their magnetic fields.

\section{The x-ray pre-cursors of the X-class flares}

\subsection{Pre-flares}

This section focuses on the pre-cursor of flares before the largest eruptions. \cite{Chifor} suggested that the x-ray pre-flares may refer to a process called tether-cutting mechanism leading to flare and filament eruption. \cite{Kim} found that a series of pre-flare events and a main flare are in causal connection being triggered by a sequential tether-cutting process. \cite{Joshi} studied a series of events consisting of three small flares followed by a major flare within two hours. They found that the pre-flares were manifestation of localised magnetic reconnection at different evolutionary states of the filament, and they played a key role in destabilising the filament leading to a major eruption. The series of pre-flares is a set of discrete, localised x-ray brightenings at intervals between 2 and 50 min before the onset of the major flare. Recently, \cite{Balazs} have shown that there is a statistical relationship between successive solar flares.

The precursors of the GOES X-class events in the Active Region NOAA 10486 have been examined by using the data of the RHESSI satellite \citep{Lin}. This active region has produced four significant X-class flares between the values of -55 and 55 degrees central meridian distance. We used the RHESSI database as a complementary dataset, which provides information about the precedents of X-flares. This satellite is also able to observe smaller and therefore more flares than the GOES satellites. Typically the RHESSI flares are mostly microflares A, B, or C of GOES class; the most frequent type of flare being GOES class B. 

\subsection{Statistical study}

 \begin{figure}
	\centerline{\includegraphics[height=0.45\textheight]{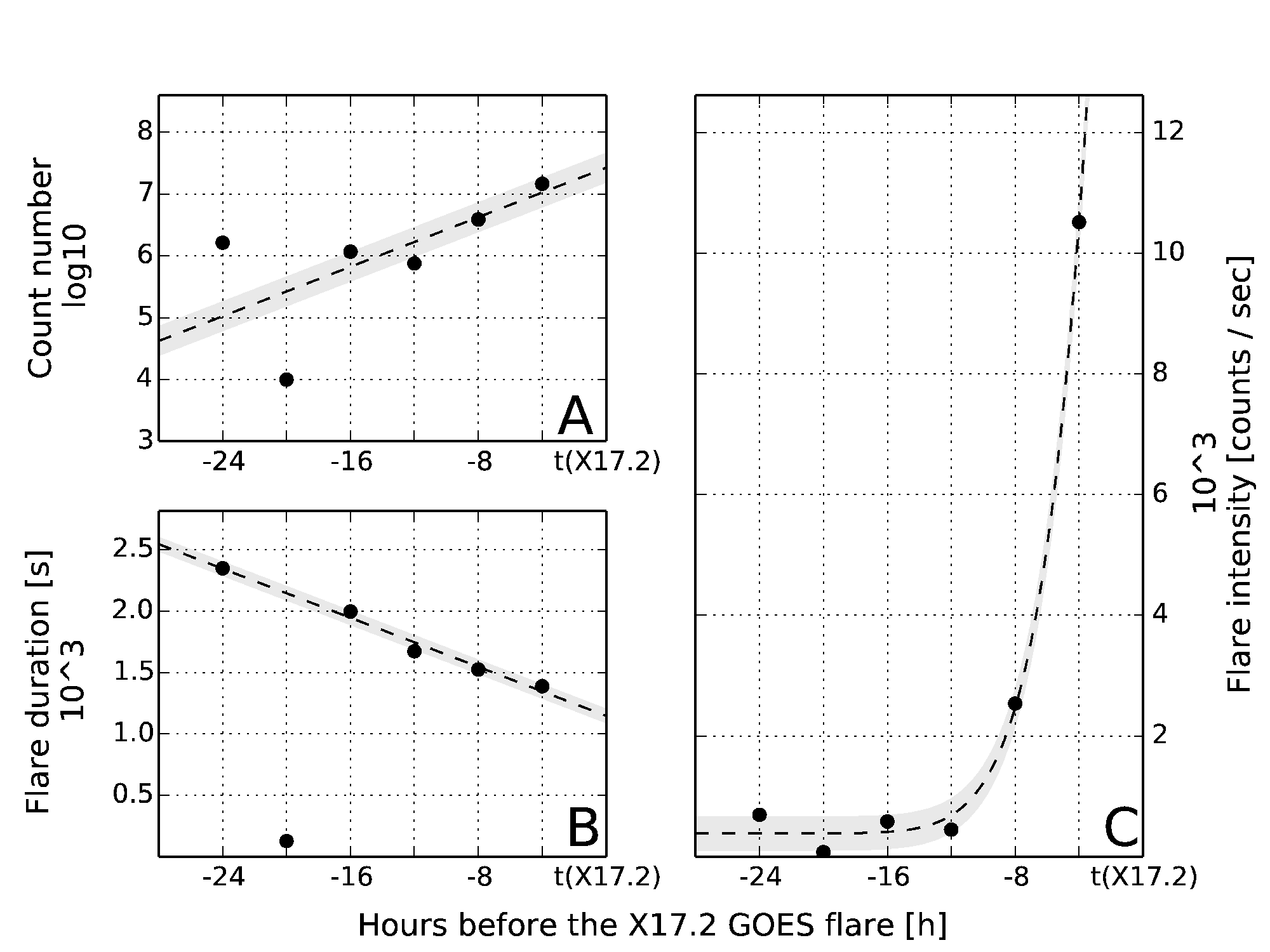}}
	\caption{Statistics of pre-cursor flares of the GOES X17.2 of NOAA 10486 based on  RHESSI x-ray events. The panels show the temporal variation of count number (A), flare duration (B) and intensity (C) by 4-hour time bins.}
	\label{precursors_X17}
\end{figure}

The RHESSI data acquired here contains information only about three X-class flare events (X17.2 on 28 October, X10 on 29 October and X8.3 on 2 November) because there were no observations in the time interval of the X1.2 flare (25 October).

In the first statistics of all the investigations shown here, which contains the data for the X17.2 flare, the events in the 24-hour interval prior to the X17.2 flare were considered. This is a sample of 18 events mostly in the energy interval of 6-25 KeV, which typically belong to the B and C GOES classes. The interval has been divided into six sub-intervals of 4 hours. Figure \ref{precursors_X17} depicts the pre-flare activity of the largest eruption (X17.2) based on RHESSI data. The dashed lines are fitted by the least-squares method in the first step, then the points above one sigma have been omitted and, in the second step, new lines have been fitted, because certain points were remarkably far from the recognizable trends. The grey belts around the lines indicate the significance level of these secondary fits.

Panel A of Figure \ref{precursors_X17} shows the temporal variation of the overall RHESSI count number. We summarized every flare event in each time bin of 4 hours. The diagram shows an increasing trend before the GOES X17.2 event. Table \ref{fitting_parameters} contains a series of the statistical properties of the trend line fitting. Panel B of Figure \ref{precursors_X17} depicts the temporal variation of the duration of the x-ray emission events, also in each time bin of 4 hours. The corresponding parameters are given in Table \ref{fitting_parameters}. The duration of x-ray emission events decreases before the X17.2 event.

\begin{table}
	 \center
	 \caption{The parameters and other properties of the fitted trend lines. The parameter x is related to one 4-hour time interval bin.}
	 
 	\begin{tabular}{@{}lrrr@{}}
		 \hline
		Panel&A&B&C\\
 		\hline
 			Model parameters&  $0.4*x+4.6$ & $-199*x+2544$  & $0.11*x^{7}+393$ \\	
			Goodness of fit ($\chi^2$)& 0.0325 &  6.03& 651 \\		
 			Standard error of&0.2239 & 50  &256 \\
 			the estimate ($\sigma$)&&& \\
		\hline
	\end{tabular}
	\label{fitting_parameters}
\end{table}

The temporal variation of flare intensity is plotted in panel 'C' of Figure \ref{precursors_X17} (right). The measure of intensity is defined as the count number divided by the flare duration. The diagram shows that the microflare intensity increases very rapidly before the X17.2 event. The increase can be represented by an exponential model function, its statistical parameters are given in Table  \ref{fitting_parameters}.

 \begin{figure}
	\centerline{\includegraphics[height=0.45\textheight]{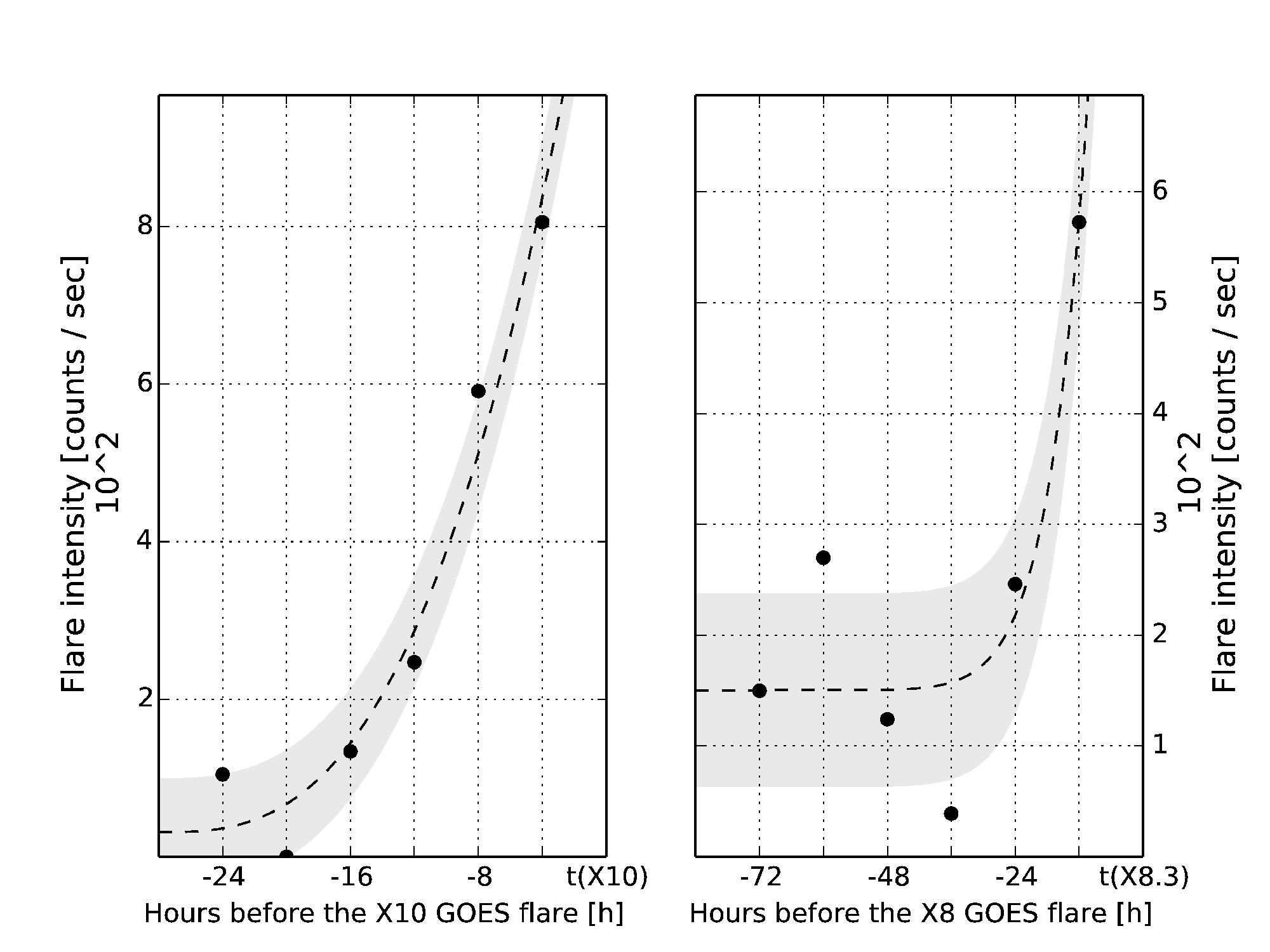}}
	\caption{Statistics of pre-cursor flares of the GOES X10 and X8.2 of NOAA10486 based on the RHESSI x-ray events. The panels show the time variation of intensity.}
	\label{precursors_X10}
\end{figure}

Similar statistics has been applied to the study of X10 and X8.3 events (Figure~\ref{precursors_X10}, left and right panels, Table \ref{fitting_parameters2}). These two series of events contain 15 and 16 microflares, respectively. The pre-flare time was also divided into 4-hour time bins before the X10 flare but the time interval before the X8.3 was as long as 72 hours, so this interval is divided to six sub-intervals of 12 hours. Both sides of the panels in Figure~\ref{precursors_X10} show exponential increase of the microflare intensities. The lengths of the intervals were not arbitrary, before the X17.2 event 24 hours were covered by observations, while before the X10 event about 24 hours passed since the previous flare, and, before the X8.3 event the development was longer, thus an interval of 72 hours was considered.

\begin{table}
	 \center
	 \caption{The parameters and other properties of the fitted trend lines of the X10 and X8 flares.
	 }
 	\begin{tabular}{@{}lrrr@{}}
		 \hline
			&precursors of X10&precursors of X8\\
 		\hline
 			Model function and parameters & $5*x^{2.8}+31$ & $0.11*x^{7}+393$ \\	
			Goodness of fit ($\chi^2$) & 216 &  192 \\		
 			Standard error of the estimate ($\sigma$) & 67  & 86 \\
		\hline
	\end{tabular}
	\label{fitting_parameters2}
\end{table}

\section{Summary}

The presented pre-defined quantities and series of phenomena allow to assess the risk and imminence of flares from pre-flare situations and developments. The following features were examined in Active Region NOAA 10486. 

1. In Paper I, the $G_{M}$ and the distance were measured between spot pairs. The parameter $WG_{M}$ defined here by Equation (2) was determined on a subset of spots at the neutral line. Its variation exhibits a similar pattern to that of described in Paper I: increase until a maximum value, then, decrease until the flare. However, an unexpected phenomenon was found: During the period of approaching and then receding motion of the two opposite polarity groups can be observed. The  $WG_{M}$ and variation of the distance is repeated four times in this very complicated active region. A further examination of the behavior of the distance and  the $WG_{M}$ may be more promising with a larger statistical sample. 

2. The series of microflares prior to the energetic flares also exhibited characteristic temporal variations. The intensity count number increased, the length of the microflares decreased nearly linearly, whereas their intensity (measured as counts/sec) increased as a steep exponential function leading to the X17.2 flare. For the flares X10 and X8 only the last type of diagram is plotted showing similar steep, accelerating rises.

3. The variation of $G_{S}$ is a quick look-type of tool providing information about an entire active region. On the other hand $WG_{M}$ is a measure of magnetism of a subgroup of sunspots of opposite polarities in the vicinity of the highest magnetic gradient in the selective area. The similarities of the two variations mean that $G_{S}$ is also useful for a quick survey and the automation of its determination is more straightforward, it does not need any subjective step.

4. The static separateness parameter defined by Equation (4) has low values under unity in the first, flare-active part of the observed time period, then it increases above unity. After a less active interval an X8 flare occurs. This parameter may be used for a quick assessment of flare probability, see e.g. \cite{Korsosb}.

The above points 1 and 2 present two faces of the pre-flare development. At photospheric level the development of the $WG_{M}$ at the location of the imminent flare is a signature of the process of the build-up of the free energy as was reported in Paper I. Its comparison to pre-flare microflares is facilitated by the grey stripes in panel A of Figure~\ref{10486grad}, Figures~\ref{precursors_X17} and Figure~\ref{precursors_X10} showing the pre-flare activities at the same time intervals. The increasing gradient is accompanied by a growing microflare activity until the major flare. This relationship is further supporting the usability of the $WG_{M}$ parameter. \cite{Jakimiec} found that the x-ray microflares and the subsequent major flare are co-located in a statistically significant number of cases. \cite{Zuccarello} carried out a thorough analysis of the events preceding the X17.2 flare and concluded that the filament destabilisation two hours before the flare onset released a domino process leading to the flare. Their study presents more details about the causal connections of the flare than ours but the above described methods seem to be more suitable for forecast because of the longer temporal range. The proposed parameters and presented processes seem to be useful tools for flare predictive activities used simultaneously in a parallel way.

\section {Acknowledgment}
The research leading to these results has received funding from the European Community's Seventh Framework Programme (FP7/2007-2013) under grant agreement eHEROES (project No. 284461). This research has made use of SunPy, an open-source and free community-developed solar data analysis package written in Python \citep{Mumford} . M. B. K. is grateful to Science and Technology Facilities Council (STFC) UK for the financial support received.

\end{document}